\documentclass[pra,aps,twocolumn,epsfig,nofootinbib]{revtex4}

\usepackage{amsmath}
\usepackage{amssymb}
\usepackage[dvips]{graphicx}%\usepackage{graphicx}% Include figure files
\usepackage{dcolumn}% Align table columns on decimal point
\usepackage{bm}% bold math
\usepackage[colorlinks=false,dvipdfm]{hyperref} % citecolor=blue 文件内部引用
\usepackage{setspace}
\usepackage{amsfonts}
\usepackage{rotating}
\usepackage{makeidx}
\usepackage{amsthm}
\usepackage{bbold}
\usepackage{float}

\newcommand{\tr}{\mbox{Tr}}

\newcommand{\Diag}{\mbox{Diag}}
\newcommand{\sgn}{\mbox{sgn}}

\begin{document}
%\begin{CJK*}{GB}{}

\title{Preparation of quantum correlations assisted by a steering Maxwell demon\footnote{Phys. Rev. E 106, 014119 (2022)}}

\author{Gang-Gang He}
\affiliation{Department of Physics, School of Science, Tianjin University, Tianjin 300072, China}

\author{Fu-Lin Zhang}
\email[Corresponding author: ]{flzhang@tju.edu.cn}
\affiliation{Department of Physics, School of Science, Tianjin University, Tianjin 300072, China}

%Although the no-signalling theorem prevents Ella
%from affecting the reduced state of Bob’s system B from
%distance [7, 8], she can remotely steer it into different
%ensembles by performing different measurements on her
%own system A

\date{\today}

\begin{abstract}
A Maxwell demon can reduce the entropy of a quantum system by performing measurements on its environment.
%This avoids the disturbance to average energy of the system.
The nonsignaling theorem prevents the demon from affecting the average state of the system.
We study the preparations of  quantum correlations from
a system qubit
%a thermal qubit
and an auxiliary qubit, assisted by a demon who obtains information of  the system qubit from measurements on its environment.
The demon can affect the postmeasured states of system by choosing different measurements, which establishes the relationships between quantum steering and other correlations in the thermodynamic framework.
We present the optimal protocols for creating mutual information, entanglement and Bell-nonlocality.
%,the maximums of which depend monotonically upon the extractable work in a similar process.
These maximal correlations are found to relate exactly to the steerable boundary of the system-environment state with maximally mixed marginals.
We also present upper bounds of the prepared correlations by utilizing classical environment-system correlation,
which can be regarded as steering-type inequalities bounding the correlations created with the aid of classical demons.
\end{abstract}
%
%LHS两句变短些
%加上最优化和充要条件
%

	
 \maketitle
	
\section{Introduction}	

The connection between thermodynamics and information provides a different angle of view to understand the physical world.
In the history of this topic, the Maxwell's demon, first introduced by Maxwell in 1871 \cite{MaxwellD}, has played an important role.
The Maxwell demon is a creature who can reduce the entropy of a system, by observing its microstates, without performing any work on it.
Szil\'{a}rd  \cite{Z.Phys.53.840} presented a one-molecule heat engine assisted by a Maxwell demon measuring the (binary) position of the molecule.
His model showed an explicit connection between information and physics that,
one can extract work $W=k T\log 2$ from the one-molecule system at a temperature $T$ by using $1$ bit information acquired by the demon.

In the field of quantum thermodynamics \cite{Book2004,Book2009},
many quantum versions of the  Maxwell demon and Szil\'{a}rd engine have been presented,
to investigate the role of quantumness in thermodynamics and  the  interplay between quantum information and thermodynamics \cite{RMP2009MaxwellD,discorddemons,NJP2017Szilard,PRL2017Demon,PRL2019NonequilibriumDemon,PRL2019,PRL2022Ren,PRL2011QSzilard,PRL2013Engine,NC2015,PRL.124.100603}.
The definitions of these models rely on the division between the quantum and classical worlds.
For instance, in Zurek's division \cite{discorddemons},
 a quantum demon is the one who
can perform global measurements on composite systems, while a  classical demon is  local.
%%%
On the other hand, quantum correlations in thermodynamics have gotten a lot of attention, as they are the most profound quantum features and  deeply connected to quantum information.
The thermodynamic cost and fundamental limitations for preparation of quantum correlations were studied under different  conditions \cite{Huber_2015,PhysRevE.100.012147}.
The correlations in turn can be used to enhance the extraction of work \cite{PRL2002ThermodynamicalCorrelations,PhysRevX.5.041011,PhysRevE.93.052140,PhysRevA.99.052320,PRL2019,discorddemons,francica2017daemonic,PhysRevLett.121.120602,PhysRevLett.122.130601,PRL2022Ren}.
The Maxwell demons and Szil\'{a}rd engines often played key roles in these works,
 such as the studies of work deficit \cite{PRL2002ThermodynamicalCorrelations}, discord  \cite{discorddemons,francica2017daemonic,PhysRevLett.121.120602,PhysRevLett.122.130601} and steering heat engines \cite{PRL2019,PRL2022Ren}.

Measurements on a quantum system would in general disturb its state, and thus affect its energy.
% the expectation value of its energy.
%
This actually provides a different paradigm in quantum thermodynamics in which measurement apparatuses are used to fuel engines \cite{PRL2017Demon,PRL2019MCool}.
That is, Maxwell demons directly measuring quantum systems lack a basic feature of their classical counterparts:
\emph{acquiring information but without affecting the state}.
%energy
The difficulty was overcome in the version of quantum Szil\'{a}rd engine presented by Beyer \emph{et al.} \cite{PRL2019},
where a demon obtains the information of a system from measurements on its environment.
The average state of the system was protected by the nonsignaling theorem.
Their approach connects the thermodynamic task of work extraction with the quantum steering,
which is a kind of quantum correlation lying between Bell nonlocality and entanglement \cite{PRL2007Steering}.
Here, the term \emph{steering}, introduced by Schr\"{o}dinger\cite{SCat},
means that the demon can project the system into different states by choosing it's measurements on the environment.
The demon's ability of steering can be convincingly demonstrated, only when the postmeasured states of the system cannot be described by a local-hidden-state (LHS) model.
In this case, the state of system and environment is said to have quantum steering from the environment to the system \cite{PRL2007Steering},
and the demon is termed \emph{truly quantum} by Beyer \emph{et al.} \cite{PRL2019}.

In this work we investigate the process for creating quantum correlations from
a system qubit
%a thermal qubit
and an auxiliary qubit assisted by a Maxwell demon measuring its environment.
This is based on the  consideration that the information acquired by demon to  extract work in \cite{PRL2019}  can certainly be used for  creating correlations.
%This is based on the  consideration that the work output of the quantum engine disigned by Beyer \emph{et al.} \cite{PRL2019} could serve as the thermodynamic cost of correlations preparation.
%The information acquired by demon to hence the extractable work can certainly be directly used for  creating correlations.
%
Our  processes connect the quantum steering with other correlations in the thermodynamic framework.
%One difference from the scheme in \cite{PRL2019}  is that,
%we allow operator to optimize the joint unitary transformations between the thermal qubit and the ancilla,
%while there are only two pairs of unitaries to choose from in the work extraction of Beyer \emph{et al.} \cite{PRL2019}.
For an arbitrary set of  observables on the environment,
we show that the maximums of quantum mutual information, entanglement and Bell-nonlocality allowed  between the system and ancilla
are all monotone increasing functions of
the average length of Bloch vectors in the postmeasured states for the system qubit.
%Consequently, they depend monotonically upon the extractable work.
%%
When the dimension of the Hilbert space of the environment can be measured by the demon is $2$,
these maximal correlations are related exactly to the steerable boundary of
the system-environment state with maximally mixed marginals.
We  also present upper bounds of the prepared correlations for unsteering demons.
These can be regarded as steering-type inequalities bounding the correlations created with the aid of classical environment-system correlation.

%%%
\begin{figure}
%[H]
 \centering
 \includegraphics[width=8cm]{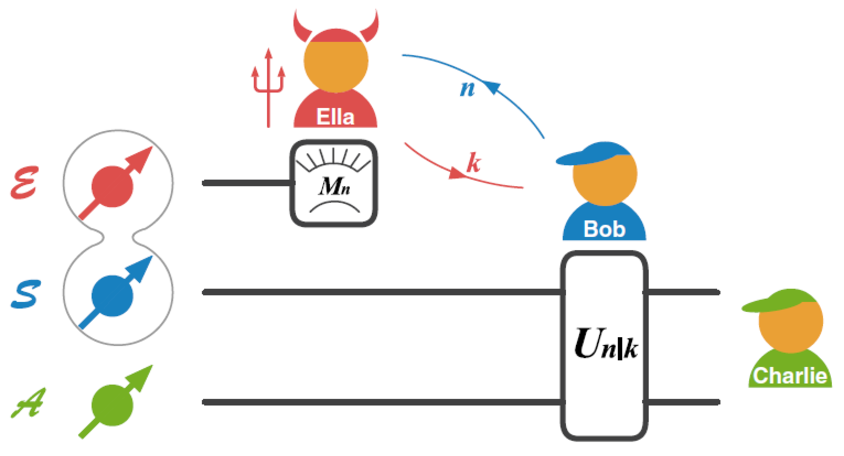}
 \caption{
 Protocol for quantum correlations preparation assisted by a  demon.
Ella and Bob share the state of $\mathcal{E}$ and $\mathcal{S}$, $\rho_{se}$.
Bob picks an observable  $M_{n}$  with a probability $q_n$ , and asks Ella to perform it on $\mathcal{E}$.
After measuring  $M_{n}$,  Ella announces to Bob her outcome $k$.
According to Ella's information, Bob performs the two-qubit unitary $U_{n|k}$ on $\mathcal{S}$ and his auxiliary qubit $\mathcal{A}$,
and sends the two qubits to Charlie.
}\label{Demon}
\end{figure}

%The paper is organized as follows.
 %In Sec. \ref{protocol},
 In the next section, we study the optimal protocols of quantum correlations preparation,
 % (including the cases without and with a demon),
 with brief introductions to the correlation measures.
In Sec. \ref{D2E}, we deal with the case that the demon performs measurements on a two-dimensional Hilbert space of the environment,
to show the advantage of a quantum demon.
Finally, a summary of our results and some outlooks are given in Sec. \ref{summary}.

\section{Correlations preparation}\label{protocol}

%

%Now, we introduce
We begin by introducing
the procedure for quantum correlations preparation assisted by the demon,
which is shown in Fig. \ref{Demon}.
%We introduce the two participants, Ella and Bob, in our protocol.
%%
Suppose that $\mathcal{S}$ is  a system qubit.
Ella is the Maxwell demon who can perform measurements on the environment $\mathcal{E}$ of $\mathcal{S}$.
Here, $\mathcal{E}$ should be understood as a part (a subsystem or a subspace of the Hilbert space) of the whole environment of $\mathcal{S}$, which Ella is  able to measure.
Bob is the operator manipulating the system qubit and an auxiliary qubit $\mathcal{A}$.
The task of Bob is to create quantum correlations from his  two qubits, in a uncorrelated initial state, by applying a global unitary on them.
In this work, we study three types of correlations between $\mathcal{S}$ and $\mathcal{A}$:
Total correlation measured by quantum mutual information $\mathcal{I}$ \cite{Book},
entanglement measured by concurrence $\mathcal{C}$ \cite{Wootters98}
and negativity $\mathcal{N}$ \cite{NEG1},
and Bell-nonlocality measured by the maximal violation of the Clauser-Horne-Shimony-Holt (CHSH) inequality  $\mathcal{B}$ \cite{RMP2014bell}.

Without loss of generality,  we set  the initial state of $\mathcal{S}$  to be
 \begin{equation}\label{Tstate}
\tau_{s}=\frac{1}{2}(\openone + \eta \sigma_z),
\end{equation}
with $ \sigma_{z} = |0\rangle \langle0|-|1\rangle \langle1|$ being the third Pauli operator.
%Suppose that
%
% a system qubit $\mathcal{S}$ is governed by the Hamiltonian
%\begin{equation}
%{H}_{s}=-\frac{\omega}{2} \sigma_{z}.
%\end{equation}
%with $ \sigma_{z} = |0\rangle \langle0|-|1\rangle \langle1|$ being the third Pauli operator.
%Its   thermal state is
%\begin{equation}\label{Tstate}
%\tau_{s}= \frac{e^{- \beta H}}{\mathcal{Z}}=\frac{1}{2}(\openone + \eta \sigma_z),
%\end{equation}
%where $\mathcal{Z }= \tr e^{- \beta H}= 2 \cosh (\beta \omega/2)$ is the partition function, $\beta=\frac{1}{k T}$ is the inverse temperature,
%and $\eta = \tanh(\beta \omega/2)$.
%%%%%%
%%%%%%
%%%%%%
%Ella is the Maxwell demon who can perform measurements on the environment $\mathcal{E}$ of $\mathcal{S}$.
%
The  whole state of $\mathcal{E}$ and $\mathcal{S}$ is $\rho_{se}$, with $\rho_s=\tr_e \rho_{se} = \tau_s$.
Suppose $\{ M_n,| n=1,2... \}$ is the set of observables Ella can measure on $\mathcal{E}$,
and $M_{n|k}$ with $k=0,1...$ denote the positive operator-valued measurement elements of $M_n$.
In each round,
Bob generates a value of $n$ with a probability $q_n$  and sends it to Ella.
Then, Ella performs $M_n$ on $\mathcal{E}$.
The probability of the outcome $k$ is $p_{n|k}= \tr[ (\openone \otimes M_{n|k}) \rho_{se} ]$,
and the corresponding collapsed state of the system is $\rho_{n|k}= \tr_e [ (\openone \otimes M_{n|k}) \rho_{se} ]/p_{n|k}$.
Each $M_n$ leads to a decomposition of the initial state as
% thermal state as
$\tau_s = \sum_{k} p_{n|k} \rho_{n|k}$.
Ella informs Bob of her outcome.
According to the outcome, Bob performs a global unitary $U_{n|k}$ on $\mathcal{S}$ and the auxiliary qubit $\mathcal{A}$ in his hands,
which are in the  initial state $\tau_s \otimes \rho_{a}$.
%Then,
% The final state of  $\mathcal{S}$ and $\mathcal{A}$ prepared by Bob with the aid of Ella is
% \begin{equation}\label{xisa}
%\xi_{s a}=\sum_{n,k} q_n  p_{n|k} \left( U_{n|k}  \rho_{n|k} \otimes \rho_{a}  {U_{n|k}}^{\dag} \right).
%\end{equation}
%Here,
We assume that another observer, Charlie, receiving the two-qubit state prepared by Bob,
knows in advance  the details of the procedure but is ignorant of the values of $n$ and $k$ in a specific run.
Therefore, the final state of  $\mathcal{S}$ and $\mathcal{A}$ received by Charlie is
% prepared by Bob with the aid of Ella is
 \begin{equation}\label{xisa}
\xi_{s a}=\sum_{n,k} q_n  p_{n|k} \left( U_{n|k}  \rho_{n|k} \otimes \rho_{a}  {U_{n|k}}^{\dag} \right).
\end{equation}

In this work, we study three types of correlations between $\mathcal{S}$ and $\mathcal{A}$:
total correlation measured by quantum mutual information $\mathcal{I}$ \cite{Book},
entanglement measured by concurrence $\mathcal{C}$ \cite{Wootters98}
and negativity $\mathcal{N}$ \cite{NEG1},
and Bell-nonlocality measured by the maximal violation of the CHSH inequality  $\mathcal{B}$ \cite{RMP2014bell}.
In the following, we derive the optimal  protocols, which maximize these correlations in $\xi_{s a}$.

\subsection{Without the demon}

Let us begin with the case without the help of Ella as a preview.
We consider an  arbitrary initial state of the system qubit
\begin{equation}
\rho_{s}=\frac{1}{2}\left(\mathbb{1}+\vec{r}\cdot\vec{\sigma}\right),
\end{equation}
 with the Bloch vector $|\vec{r}|=r\in [0,1]$ and $\vec{\sigma}$ being the vector of Pauli matrices.
 Specifically, in the following, we show all the maximal correlations of
\begin{equation}\label{zetasa}
\zeta_{sa}= U \rho_{s}\otimes \rho_a  U^{\dag} ,
\end{equation}
 among all the global unitaries $U$ and initial states of  $\mathcal{A}$ $\rho_a$,  are monotonic increasing functions of $r$.
 To reach these maximums, the initial state of  $\mathcal{A}$ can be chosen as $\rho_a= |0\rangle \langle 0|$,
 and the global unitaries $U$ can be  implemented in two steps:
(1) a local unitary diagonalizing $\rho_{s}$ into $\frac{1}{2}\left(\mathbb{1}+r \sigma _z  \right)$;
(2) a global unitary $U_0$ such that $U_0 |00\rangle =|\psi_+\rangle $  and  $U_0 |10\rangle =|10\rangle $ for entanglement while
 $U_0 |10\rangle =|\psi_-\rangle $ for mutual information  and Bell-nonlocality.
 Here $|\psi_{\pm}\rangle =(|00\rangle \pm |11\rangle)/\sqrt{2} $ are two of the Bell states.

%$\ $

%
\emph{Mutual information.--}
The quantum mutual information of a bipartite state $\rho_{\alpha \beta}$ is defined as
\begin{equation}
%\mathcal{I}(\rho_{\alpha \beta})=\mathcal{S} (\rho_{\alpha })+\mathcal{S} (\rho_{\beta }) - \mathcal{S} (\rho_{\alpha \beta}),
\mathcal{I}(\rho_{\alpha \beta})=\mathcal{H} (\rho_{\alpha })+\mathcal{H} (\rho_{\beta }) - \mathcal{H} (\rho_{\alpha \beta}),
\end{equation}
where $\rho_{\alpha } =\tr _{\beta} \rho_{\alpha \beta} $ and $\rho_{\beta} =\tr _{\alpha} \rho_{\alpha \beta} $
 are the reduced states of subsystems $\alpha$ and $\beta$, respectively,
 and $\mathcal{H} (\rho) =-\tr (\rho \log \rho)$ is the von Neumann entropy of $\rho$.
It measures the total correlation between the two subsystems, which does not distinguish classical correlation from quantum one \cite{RMP2012Vedral}.

The total entropy of the whole state of $\mathcal{S}$ and $\mathcal{A}$ is conserved under the global unitary $U$ in Eq. (\ref{zetasa}),
i.e., $\mathcal{H}(\zeta_{sa})= \mathcal{H}(\rho_{s} \otimes \rho_{a}) = \mathcal{H}(\rho_{s})+\mathcal{H}(\rho_{a})$.
It can be minimized by choosing $\rho_a= |0\rangle \langle 0|$ when the initial state  $\mathcal{S}$ is fixed.
The local entropies of $\zeta_{sa}$ are upper bounded by the  dimension of the two subsystems as $\mathcal{H}(\zeta_{s}) \leq \log2 $ and $\mathcal{H}(\zeta_{a}) \leq \log2$.
The two equalities hold when the eigenstates of $\zeta_{sa}$ are  maximally entangled states.
%The total entropy $\mathcal{H}(\zeta_{sa})= \mathcal{H}(\rho_{s})+\mathcal{H}(\rho_{a})$ is conserved under the global unitary $U$,
%and the local entropies are upper bounded by their dimension as $\mathcal{H}(\zeta_{s}),\mathcal{H}(\zeta_{a}) \leq \log2$.
These lead to the maximum of the mutual information as
 \begin{equation}\label{MI0}
\max_{\{U,\rho_a\} } \mathcal{I}(\zeta_{s a})=2 \log2- \mathcal{H} (\rho_{s})=2 \log2- h(r),
\end{equation}
where $h(r)= - \frac{1+r}{2} \log \frac{1+r}{2} - \frac{1-r}{2} \log \frac{1-r}{2}$.
It is
% can be achieved
by
%choosing $\rho_a= |0\rangle \langle 0|$
%and transforming
the whole state
%into
$\zeta_{s a}= \frac{1+r}{2} |\psi_+\rangle \langle \psi_+| + \frac{1-r}{2}  |\psi_-\rangle \langle \psi_-|$.
%

%$\ $

\emph{Entanglement.--}
Concurrence and negativity are the two most widely used measures of entanglement in two-qubit states.
The former
leads to a computable formula  for entanglement of formation  in the two-qubit case  \cite{Wootters98}.
The latter is closely related to  partial transpose criterion of entanglement \cite{NEG1}.
%which is necessary and sufficient in $2\otimes 2 $ and $2\otimes3$ systems while  is only sufficient for higher dimensional systems.
Both of them
do not increase on average, which is the  condition of convexity usually satisfied by known entanglement measures \cite{RevModPhys.81.865}.
Based on the convexity, one can easily draw a conclusion that the maximal concurrence and negativity can be simultaneously reached by choosing  $\rho_a= |0\rangle \langle 0|$.
Without loss of generality,
we suppose $\rho_a= p_0 |0\rangle \langle 0|+ p_1 |1\rangle \langle 1| $ with $p_{0,1} \in [0,1]$ and $p_{0} +p_{1} =1$,
%  show the computation procedure as
and only show the computation procedure for $\mathcal{C}$ as
 \begin{eqnarray}\label{Convex}
 \mathcal{C}(\zeta_{s a}) &\leq& p_0  \mathcal{C}( U \rho_{s} \! \otimes |0\rangle \langle 0|  U^{\dag} ) + p_{1}  \mathcal{C}( U \rho_{s} \! \otimes |1\rangle \langle 1|  U^{\dag} )\ \ \  \nonumber \\
 &\leq&  p_0  \max_{\{U\}} \mathcal{C}( U \rho_{s} \! \otimes |0\rangle \langle 0|  U^{\dag} ) \nonumber \\
 &\ &  + p_{1} \max_{\{U\}}  \mathcal{C}( U \rho_{s} \! \otimes |1\rangle \langle 1|  U^{\dag} ) \ \  \ \    \\
 &=&    \max_{\{U\}} \mathcal{C}( U \rho_{s}\! \otimes |0\rangle \langle 0|  U^{\dag} ).  \nonumber
\end{eqnarray}
Then, both of the measures of entanglement for the rank 2 state $U \rho_{s}\! \otimes |0\rangle \langle 0|  U^{\dag}$ can be maximized among global unitaries $U$
by
%Both of the measures of entanglement are maximized by the state
$\zeta_{s a}=U \rho_{s}\! \otimes |0\rangle \langle 0|  U^{\dag}= \frac{1+r}{2} |\psi_+\rangle \langle \psi_+| + \frac{1-r}{2}  |10\rangle \langle10 |$  \cite{PRA.62.022310,PRA2001MEMS},
and
 \begin{eqnarray}\label{CN0}
&&
\max_{\{U,\rho_a\} } \mathcal{C}(\zeta_{s a})=\frac{1}{2} (1+r),\\
&&\max_{\{U,\rho_a\} } \mathcal{N}(\zeta_{s a})=\frac{1}{2} \left[\sqrt{2(1+r^2)} -1+r\right].
\end{eqnarray}
It is  interesting that,
the final state $\zeta_{s a}$ above  has the minimal negativity for a fixed concurrence \cite{verstraete2001comparison}.

%
%

%$\ $

\emph{Bell-nonlocality.--}
Bell-nonlocality exists in the states whose outcomes of local measurements do not admit by any local-hidden-variable models,
which can be witnessed  by the violation of Bell-type inequalities.
We adopt the maximal quantum violation of the CHSH inequality, $\mathcal{B}$, as the degree of Bell-nonlocality for two-qubit systems.
For a two-qubit state $\rho$ with spin correlation matrix $T$,
 whose elements $T_{ij}=\tr(\sigma_i \otimes \sigma_j \rho)$ ($i,j=1,2,3$),
 $\mathcal{B}(\rho)=2 \sqrt{t^2_1 + t^2_2}$ \cite{horodecki1995violating}.
 Here,  $t^2_1$ and $ t^2_2 \in[0,1]$  are the two largest eigenvalues of $T^T T$.
The amount $\mathcal{B}(\rho) >2$ demonstrates the Bell-nonlocality of $\rho$.

In the region of nonlocality, for a fixed linear entropy $\mathcal{S}_{L}(\rho)=\frac{4}{3}(1-\tr \rho^2)$,
$\mathcal{B}(\rho) $ is maximized by the rank $2$ states mixed by any two of the Bell states \cite{ZhangPRA2016}.
And, the maximal $\mathcal{B}$ decreases with $\mathcal{S}_{L}$.
Therefore, for a given $\rho_s$,
the maximal  Bell-nonlocality that can be created is
 \begin{equation}\label{B0}
\max_{\{U,\rho_a\} } \mathcal{B}(\zeta_{s a})=2\sqrt{(1+r^2)},
\end{equation}
which is reached by
$\rho_a= |0\rangle \langle 0|$
and
$\zeta_{s a}= \frac{1+r}{2} |\psi_+\rangle \langle \psi_+| + \frac{1-r}{2}  |\psi_-\rangle \langle \psi_-|$.

\subsection{Assisted by the demon}

%The procedure above can also act as the  scenario of work extraction,  in which  $\mathcal{A}$ plays the role of a work storage system \cite{PRL2019}.
%The maximum of average extractable work can be derived by optimizing each of the terms $U_{n|k}  \rho_{n|k} \otimes \rho_{a}  {U_{n|k}}^{\dag}$ individually.
%In addtion, the result of  an global unitary, conserving total energy of  $\mathcal{S}$ and $\mathcal{A}$, on $\mathcal{S}$ is always equivalent to a local rotation of its Bloch vector \cite{PRX2016U}.
%Hence, the maximal work can be directly obtianed by rotating all the Bloch vectors $\vec{r}_{n|k} $ of the postmeasured states $\rho_{n|k} $ into Z axis, as
%\begin{equation}\label{work}
%\max_{\{U_{n|k},\rho_a\} } W(\xi_{s a})=\frac{\omega}{2}(\bar{r} -\eta),
%\end{equation}
% where $\bar{r}= \sum_{n,k} q_n  p_{n|k} r_{n|k}$ is the average length of $\vec{r}_{n|k} $.
%
%In the following, we show all the maximal prepared quantum correlations of $\xi_{s a}$ are monotonic increasing functions of $\bar{r}$.

Now, we allow Ella to participate.
For fixed $\{ q_n \}$ and  $\{ M_n \}$,  we define the average length of the Bloch vectors  $\vec{r}_{n|k} $ of  $\rho_{n|k} $ in Eq. (\ref{xisa}) as $\bar{r}= \sum_{n,k} q_n  p_{n|k} r_{n|k}$.
The maximal quantum correlations of $\xi_{s a}$ can be obtained by  replacing $r$ in Eqs. (\ref{MI0}) and (\ref{CN0})-(\ref{B0}) with $\bar{r}$.
% are monotonic increasing functions of $\bar{r}$,
%where $\bar{r}= \sum_{n,k} q_n  p_{n|k} r_{n|k}$ is the average length of the Bloch vectors of  $\rho_{n|k} $ in (\ref{xisa}).
%$\vec{r}_{n|k} $.
%%
These maximums can be reached  by optimizing each of the terms  $U_{n|k}  \rho_{n|k} \otimes \rho_{a}  {U_{n|k}}^{\dag}$
in the ways given in the above subsection.
Namely, one chooses $\rho_{a}=|0\rangle \langle0|$ and $U_{n|k}$,
and transforms $\rho_{n|k} \otimes |0\rangle \langle0|$ into the mixtures of $|\psi_+\rangle$ and $|\psi_-\rangle$ for maximal $\mathcal{I}$ and $\mathcal{B}$ while into the mixtures of $|\psi_+\rangle$ and $|10\rangle$ for maximal entanglement.
These are nontrivial, as the quantum correlations are nonlinear functions of the state.
%, while the inner energy (or  work) is  linear.
 %%
The details are in the following.

%$\ $

\emph{Mutual information.--}
The minimum of the entropy for $\xi_{s a}$ can be achieved by setting:
(1) $\rho_{a}=|0\rangle \langle0|$;
(2)  the elements are diagonal in the same set of  basis as $U_{n|k}  \rho_{n|k} \otimes |0\rangle \langle0| {U_{n|k}}^{\dag}=\sum_{i} \lambda_{{n|k},i}   |\phi_i \rangle \langle \phi_i |$ with $\lambda_{{n|k},0}  \geq \lambda_{{n|k},1} $.
The first point can be easily proved by using the concavity property of the von Neumann entropy \cite{Book},
as the calculation of concurrence in Eq. (\ref{Convex}).
Here, we omit the procedure for brevity.
The second point can be derived based on the Lemma 1 in Appendix \ref{Lemma1}.
Namely, we set $X_{n|k}= q_n  p_{n|k}  (\openone -  U_{n|k}  \rho_{n|k} \otimes |0\rangle \langle0|  {U_{n|k}}^{\dag}  )$
 and $X=\sum_{n,k} X_{n|k} = \openone - \xi_{sa}$.
The von Neumann entropy for the two-qubit state can be written as
\begin{equation}
\mathcal{H}(\xi_{sa}) = 3- \sum_{n=2}^{+\infty} \frac{1}{n(n-1)} \tr(X^n).
\end{equation}
All the terms $\tr(X^n)$ can be simultaneously maximized by the above condition (2).

Then, the entropies of two reduced states of $\xi_{s a}$ can be maximized by transforming $|\phi_0 \rangle$ and $|\phi_1 \rangle$ into the two Bell states $|\phi_+ \rangle$ and $|\phi_- \rangle$  without affecting entropy for $\xi_{s a}$.
These lead to the maximal  mutual information as
 \begin{equation}
\max_{\{U_{n|k},\rho_a\} } \mathcal{I}(\xi_{s a})=2 \log2- h(\bar{r}).
\end{equation}

\emph{Entanglement.--}
One can still use the procedure in Eq. (\ref{Convex}) to restrict in the initial state of $\mathcal{A}$ as $\rho_{a}=|0\rangle \langle0|$.
The convexity of concurrence  further gives
 \begin{equation}
 \mathcal{C}(\xi_{s a}) \leq  \sum_{n,k} q_n  p_{n|k}   \mathcal{C}( U_{n|k}  \rho_{n|k} \otimes |0\rangle \langle0|  {U_{n|k}}^{\dag} ).
\end{equation}
When the states $U_{n|k}  \rho_{n|k} \otimes |0\rangle \langle0|  {U_{n|k}}^{\dag}  =   \frac{1+r_{n|k} }{2} |\psi_+\rangle \langle \psi_+| + \frac{1-r_{n|k} }{2}  |10\rangle \langle 10|$,
each concurrence of the right-hand side reaches its maximum,
and meanwhile, the equality holds.
Then, the maximum of concurrence is given by
 \begin{eqnarray}
\max_{\{U_{n|k},\rho_a\} } \mathcal{C}(\xi_{s a})=\frac{1}{2} (1+\bar{r}).
\end{eqnarray}
These choices simultaneously maximize the negativity as
 \begin{eqnarray}
\max_{\{U_{n|k},\rho_a\} } \mathcal{N}(\xi_{s a})=\frac{1}{2} \left[\sqrt{2(1+\bar{r}^2)} -1+ \bar{r} \right].
\end{eqnarray}
We give the details in Appendix \ref{MaxN}.

\emph{Bell-nonlocality.--}
The optimization of the Bell-nonlocality measured by $\mathcal{B}$ can be solved by  minimizing the linear entropy again.
The minimum of $\mathcal{S}_L(\xi_{s a})$ is achieved under the same two conditions above for the von Neumann entropy,
which can be proved by using a similar procedure.
Then, the state $\xi_{s a}$ is rank $2$.
By transforming its eigenstates into   $|\phi_+ \rangle$ and $|\phi_- \rangle$,
Bob obtains the maximal Bell-nonlocality assisted by the demon as
 \begin{equation}
\max_{\{U,\rho_a\} } \mathcal{B}(\zeta_{s a})=2\sqrt{(1+{\bar{r}}^2)}.
\end{equation}

\subsection{Optimal measurements of the demon}
%%%%

It would be interesting to find out Ella's optimal measurements, which maximize $\bar{r}$.
Because of the convex form of $\bar{r}= \sum_{n,k} q_n  p_{n|k} r_{n|k}$, the maximum of $\bar{r}$ occurs at $q_n  \in \{0, 1\}$.
That is, the maximum is equal to  $\max_{\{n\}} \sum_{k}   p_{n|k} r_{n|k} $.

%%
%For a pure state of  $\mathcal{E}$ and $\mathcal{S}$,
When $\rho_{se}$ is pure,
any von Neumann measurement on $\mathcal{E}$ can project $\mathcal{S}$ into a pure local state,
and thereby $\bar{r}=1$ reaches its maximum.
However, for mixed states between  $\mathcal{E}$ and $\mathcal{S}$,
it is difficult to uniformly optimize Ella's measurement,
even if for the two-qubit case (i.e., $\mathcal{E}$ is two-dimensional) under local von Neumann measurements.
We give some results for this simple case below.
%%%%
%%%%
%%%%

A general form of the two-qubit state,
whose reduce state of  $\mathcal{S}$ part is $\tau_s$ in Eq. (\ref{Tstate}),
 can be expressed as
%the initial state for $\mathcal{S}$ and $\mathcal{E}$ can be expressed as
\begin{equation}\label{rhose}
\rho_{se}=\frac{1}{4}(\mathbb{1}+\eta\sigma_{z}\otimes\mathbb{1}+\mathbb{1}\otimes\vec{\sigma}\cdot\vec{b}+\sum_{ij} T_{ij}\sigma_{i}\otimes\sigma_{j}),
\end{equation}
where $\vec{b}$ is the Bloch vector on Ella's side and  $T$ is the  $3\times3$ spin correlation matrix.
An observable of Ella can be labeled by a unit vector as  $M_{\vec{n}} = \vec{n} \cdot \vec{\sigma}$,
and its elements are two projectors
\begin{eqnarray}
M_{\vec{n}|k}=\frac{1}{2}(\openone+ k \vec{n} \cdot \vec{\sigma})
\end{eqnarray}
corresponding to the outcomes $k=\pm1$.
After her measurements, the system qubit is left in the unnormalized state
\begin{eqnarray}\label{AssemC}
\rho_{\vec{n}|k}=\frac{1}{2}\left[\openone + \frac{\eta \sigma_z + k (T\vec{n}) \cdot \vec{\sigma}}{ 1 + k \vec{b}\cdot \vec{n}} \right] ,
\end{eqnarray}
with the measurement probability $p_{\vec{n}|k} =\frac{1}{2}(1 + k \vec{b}\cdot \vec{n})$.
Then,
 one can obtain the the maximal $\bar{r}$ as
\begin{eqnarray}\label{maxr}
\bar{r}= \frac{1}{2}  \max_{\{ \vec{n} \}} \bigr(|\vec{\eta} +  T \vec{n}|+|\vec{\eta} -  T \vec{n}|\bigr),
\end{eqnarray}
where $\vec{\eta}=(0,0,\eta)$.

When the Bloch vecter $\eta =0$,
it is obvious that the maximum of $\bar{r}$ is the maximal absolute value of the eigenvalues of  $T$.
The optimal $\vec{n}$ is the corresponding  eigenvector.
However, when  $\eta \neq 0$, the optimal solution depends on  both $\eta$  and $T$.
We adopt the state in the
case study of Beyer \emph{et al.} \cite{PRL2019}  as an example,
%to show the variation caused by the difference, as well as the advantage of a quantum correlated state.
%%%
%
which is
\begin{equation}\label{example}
\rho(p,\eta)=p  |\Psi\rangle  \langle\Psi |  +(1-p)\rho_{\mathrm{cl}},
\end{equation}
% %
with the two components
\begin{equation}\label{psi}
 |\Psi\rangle=\sqrt{\frac{1+\eta}{2}}|00\rangle+\sqrt{\frac{1-\eta}{2}}|11\rangle,
\end{equation}
 and
\begin{equation}\label{rhocl}
 \rho_{\mathrm{cl}}=\frac{1+\eta}{2}|0\rangle\langle0|\otimes|0\rangle\langle0|+\frac{1-\eta}{2}|1\rangle\langle1|\otimes|1\rangle\langle1|.
\end{equation}
 Its reduced state of the system qubit is $\tau_s$ in (\ref{Tstate}), and correlation matrix $T=\Diag \{p\sqrt{1-\eta^2},-p\sqrt{1-\eta^2},1 \} $.
Ella's  optimal  measurement is $M_{\vec{n}}=\sigma_z$, which reaches the maximal $\bar{r}=1$.
 Replacing the projectors $|0\rangle\langle0|$ and $|1\rangle\langle1|$ of $\mathcal{E}$ in $ \rho_{\mathrm{cl}}$ with $|+\rangle\langle+|$ and $|-\rangle\langle-|$,
 one can directly find that $\bar{r}$ can reaches $1$ only if  $p=0$ or $1$,
 and the optimal  measurement  is no longer $\sigma_z$ when $p\neq0$.

%%普遍混态求不出
%%2qubit 正交测量
%% Bell对角态时 是本征值问题
%%给出一个例子来

\section{Classical and quantum demons}\label{D2E}

We  continue to consider the simple case of a two-dimensional $\mathcal{E}$  under von Neumann measurements,
to succinctly show the different effects between a quantum demon and a classical one.
%%
%%%
The maximal prepared  correlations are monotonic increasing functions of  the average length of the Bloch vectors in  postmeasured states of the system qubit.
Therefore, the length change of the  Bloch vector,
 $\Delta r=\bar{r}-\eta$,
measures the quantum correlations  enhanced  by the participation of the demon.
%
%It also indicates the corresponding relationships between the correlations and extractable work in (\ref{work}).
In this part, we adopt the increase in the created entanglement
\begin{equation}
\Delta \mathcal{C}=\frac{1}{2}(\bar{r}-\eta),
\end{equation}
which is proportional to  $\Delta r$, as a figure of merit of the demon.

The operator, Bob, is assumed to know the form of state $\rho_{se}$ to choose Ella's measurements and perform his optimal operations.
%the optimal operations.
The triple $(\eta,T,\vec{b})$ is his \emph{a priori} knowledge of the system qubit and its environment.
Ella's measurements and outcomes convert these information to the system $\mathcal{S}$, as shown in the form of $\rho_{\vec{n}|k}$ in Eq. (\ref{AssemC}),
which leads to the enhancement of created correlations.

The case with the Bloch vectors $\eta=0$ and $\vec{b}=0$ is notable,
 in demonstrating the relationship between the created correlations and the quantumness in $\rho_{se}$.
That is, the system  $\mathcal{S}$ is in the high temperature limit $T\rightarrow +\infty$ and the local state of $\mathcal{E}$ is completely unknown.
Such a $\rho_{se}$ is equivalent to a Bell diagonal state \cite{horodecki1996information,dakic2010necessary,PRA.99.062314} under some local unitary transformations.
When the measurement directions $\vec{n}$ are completely randomized on the unit sphere,
the entanglement enhanced by Ella is given by
\begin{equation}
\Delta\mathcal{C}  =   \int \frac{1}{8 \pi}  |T \vec{n}| d \vec{n},
\end{equation}
where the integral is over the unit sphere and $d \vec{n}$ is the surface element.
A critical value,
  $\Delta\mathcal{C}_c = 1/4 $,
corresponds exactly to the necessary and sufficient condition for steerability of the state $\rho_{se}$ \cite{PRA.99.062314,JOSAB2015Steering,EPL2016Tstate}.
Consequently, the amounts of  the quantum correlations prepared with the assistance of the demon can serve as criterions and measures
 for steerability of the state $\rho_{se}$ in this case.

For a general two-qubit $\rho_{se}$, we focus on the situation with two observables.
This is the minimum number of observables to show the advantage of a quantum Maxwell demon.
%, she should have at least two observables.
%This is because
%The quantumness, characterized by Ella's ability of steering, is demonstrated by the difference between the postmeasured states of $\mathcal{S}$  corresponding different local measurements on $\mathcal{E}$.
%%%
%The optimal results with one measurement $M_{\vec{n}} $ can always be achieved by the classically correlated state
%\begin{equation}
%\rho_{se}=\frac{1+\eta}{2}|0\rangle \langle  0 | \otimes M_{\vec{n}}_{+1}+\frac{1-\eta}{2}|1\rangle \langle  1 | \otimes M_{\vec{n}}_{-1}.
%\end{equation}
% The state $\rho_{\mathrm{cl}}$ in (\ref{rhocl}) belongs to such form.
%In this case, the average length of the Bloch vectors of postmeasured states achieves $\bar{r}=1$.
%
%We now revisit
%We now focus on
%the situation with a general $\rho_{se}$ and two observables to reveal the quantumness of the demon.
%%%%
In addition, we assume that Bob requires the two measurement directions to satisfy $T\vec{n}_{1} \perp T\vec{n}_{2}$,
 according to his a \emph{priori} knowledge of the matrix $T$.
That is, the changes of direction of the Bloch vector of $\mathcal{S}$ affected by Ella's measurements are perpendicular to each other.
%%%
These two points conform to the intuitive understanding from the classically correlated state $\rho_{cl}$ in Eq. (\ref{rhocl}).
It cannot be distinguished from the fully quantum state $ |\Psi\rangle$ in Eq. (\ref{psi}) by the measurement on $\sigma_z$,
which leads to the maximum $\bar{r}=1$.
A feature of $\rho_{cl}$ is that,
 corresponding to any measurement on  $\mathcal{E}$,
 the change of the Bloch vector of $\mathcal{S}$ is along the $z$ axis.
%
%The maximum, $\bar{r}=1$, can be reached by the measurement on $\sigma_z$,
% and hence it can not be distinguished from the fully quantum state $ |\Psi\rangle$ in (\ref{psi}).
%
%
%One measurement on $\sigma_z$ reaches the maximal $\bar{r}=1$
%
% is $M_{\vec{n}}=\sigma_z$, which reaches the maximal $\bar{r}=1$.
%
%这两点是符合直观的，通过观察rhocl的经典态。
%一个测量如果选择z方向
%
%
For the case with two observables onto the two general states (\ref{rhose}),
%denoted by  the two unit vectors $\vec{n}_1$ and $\vec{n}_2$,
chosen by Bob with equal probabilities,
one can directly obtain that the entanglement enhanced by Ella is
\begin{equation}\label{DeltC}
\Delta\mathcal{C}  = \frac{1}{8} \biggr( - 4\eta  + \sum_{i=1,2;k=\pm1} |\vec{\eta} + k T \vec{n}_{i}|  \biggr).
\end{equation}
%where $\vec{\eta}=(0,0,\eta)$.
%%%%
%In addition, we assume that Bob requires the two measurement directions to satisfy $T\vec{n}_{1} \perp T\vec{n}_{2}$,
%according to his a \emph{priori} knowledge of the matrix $T$.
%That is, the changes of direction of the Bloch vector of $\mathcal{S}$ affected by Ella's measurements are perpendicular to each other.
%
%The quantumness, characterized by Ella's ability of steering, is demonstrated by the difference between the postmeasured states of $\mathcal{S}$  corresponding different local measurements on $\mathcal{E}$.
%Therefore,  it is reasonable to assume that Bob requires the two measurement directions to satisfy $T\vec{n}_{1} \perp T\vec{n}_{2}$,
%according to his a \emph{priori} knowledge of the matrix $T$.
%That is, the changes of direction of the Bloch vector of $\mathcal{S}$ affected by Ella's measurements are perpendicular to each other.
%%Under this condition, the optimal two measurements of the $\rho(p,\eta)$ in Fig. \ref{figwork} to create the maximal quantum correlations are exactly $M_{\vec{n}_1} = \sigma_x $ and  $M_{\vec{n}_2}=\sigma_z$.
%%%
%%
Below we present an upper bound on the enhanced concurrence under this condition for classical (unsteerable) demons.

A LHS model admitted by a two-qubit state $\rho_{se}$ can be identified with a hidden Bloch vector $\vec{\lambda}$ with a distribution $\omega(\vec{\lambda})$,
 and
a function $f (\vec{n}, \vec{\lambda}) \in [-1,1]$ of $\vec{\lambda}$ and  the measurement direction $\vec{n}$  \cite{PRA.99.062314}.
They satisfy
\begin{align}
\begin{split}
& \! \! \! \int\omega(\vec{\lambda}) \biggr(\openone +\vec{\lambda} \cdot \vec{\sigma} \biggr)d\vec{\lambda}=\openone + \eta \sigma_z ,\\
&\! \! \!  \int\omega(\vec{\lambda})f(\vec{n},\vec{\lambda})  \biggr(\openone +\vec{\lambda} \cdot \vec{\sigma} \biggr) d\vec{\lambda}=(\vec{n}\cdot\vec{b} ) \openone  +(T\vec{n})  \cdot \vec{\sigma},  \\
\end{split}
\end{align}
 where the integral is over the Bloch sphere and $d \vec{\lambda}$ is the surface element.
We denote $T\vec{n}_{1}=\vec{\alpha}_{1}$ and $T\vec{n}_{2}= \vec{\alpha}_{2}$.
The average length of Bloch vectors in Eq. (\ref{DeltC})  satisfies
%%%
\begin{equation}\label{rbar}
\bar{r}\leq \frac{1}{2}\left(\sqrt{\eta^{2}+\alpha^{2}_{1}}+\sqrt{\eta^{2}+\alpha^{2}_{2}}\right),
\end{equation}
and the two changes
\begin{equation}\label{alpha}
\alpha^{2}_{i}=\int\omega(\vec{\lambda}) f(\vec{n}_{i},\vec{\lambda})(\vec{\lambda}\cdot\vec{\alpha}_{i})d\vec{\lambda},
 %%%
\end{equation}
with $i=1,2$.

The amount of $\alpha^{2}_{i}$ is upper bounded by $f(\vec{n}_{i},\vec{\lambda})=\sgn (\vec{\lambda}\cdot \vec{\alpha}_{i})$, where $\sgn $ is the sign function.
%%%
Without loss of generality, we choose $\vec{\alpha}_{1}= {\alpha}_{1} \vec{i} $ and $\vec{\alpha}_{2}= {\alpha}_{2} \vec{j} $,
where $\vec{i} $ and  $\vec{j} $ are  unit vectors in the  $x$ and  $y$ directions, respectively.
%%%
%Set the unit vectors $\vec{e}_{1} = \vec{\alpha}_{1}/{\alpha}_{1} $,  $\vec{e}_{2} = \vec{\alpha}_{2}/{\alpha}_{2} $ and
%$\vec{e}_{3} =\vec{e}_{1}  \times \vec{e}_{2}$.
%Under such a set of bases,
When $f(\vec{n}_{i},\vec{\lambda})=\sgn (\vec{\lambda}\cdot \vec{\alpha}_{i})$,
 the  integral (\ref{alpha})  is invariant under the  inversions of  the distribution  that
\begin{eqnarray}
&& \omega(\lambda_1,\lambda_2,\lambda_3)  \rightarrow \omega(-\lambda_1,\lambda_2,\lambda_3), \nonumber\\
&& \omega(\lambda_1,\lambda_2,\lambda_3)  \rightarrow \omega(\lambda_1,-\lambda_2,\lambda_3) . \nonumber
\end{eqnarray}
%The  integral  above with $f(\vec{n}_{i},\vec{\lambda})=\sgn (\vec{\lambda}\cdot \vec{\alpha}_{i})$ is invariant to $\omega(\lambda_1,\lambda_2,\lambda_3)  \rightarrow \omega(-\lambda_1,\lambda_2,\lambda_3)  $ and $\omega(\lambda_1,\lambda_2,\lambda_3)  \rightarrow \omega(\lambda_1,-\lambda_2,\lambda_3)  $.
Consequently,
the maximum of  the right-hand side of Eq. (\ref{rbar})
can always been found  among the distributions which are invariant under the above two  inversions.
%$\omega(\lambda_1,\lambda_2,\lambda_3) = \omega(-\lambda_1,\lambda_2,\lambda_3)  $ and $\omega(\lambda_1,\lambda_2,\lambda_3) = \omega(\lambda_1,-\lambda_2,\lambda_3)  $.
%
%
Then,   $\alpha^{2}_{i}$ are determined by $\vec{q}=\int_{\lambda_1\geq0,\lambda_2\geq0}\omega(\vec{\lambda})\vec{\lambda} d\vec{\lambda}$.
An upper bound of the average length of Bloch vectors is given by $\vec{q} = (\frac{1}{4 \sqrt{2}},\frac{1}{4 \sqrt{2}},0)$,
which is
$
\bar{r}\leq\sqrt{\frac{1}{2}+\eta^2}.
$
 %%%%
  %%%%
Considering that a physical Bloch vector is not longer than $1$,
one can conclude that the entanglement enhanced by a classical demon is bounded by
 %%%%
\begin{align}\label{DeltCB}
\begin{split}
\Delta\mathcal{C}_{\mathrm{cl}}  \leq \frac{1}{2}\left(-\eta+ \min\biggr\{\sqrt{\eta^{2}+\frac{1}{2}},1\biggr\}\right).
\end{split}
\end{align}

\begin{figure}
%[H]
 \centering
 \includegraphics[height=5cm]{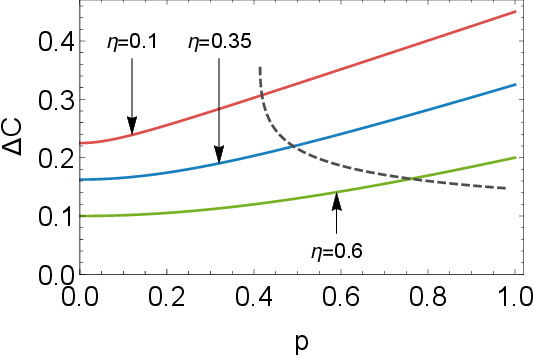}
 \caption{
Enhanced concurrence of the state $\rho(p,\eta)$ under the measurements on $M_{\vec{n}_1} = \sigma_x $ and  $M_{\vec{n}_2}=\sigma_z$ with equal
probabilities.
The solid lines show  enhanced concurrence $\Delta \mathcal{C}$ with the parameter  $\eta=0.1$, $\eta=0.35$, and $\eta=0.6$  from top to bottom.
% Extractable work (enhanced concurrence)  in Fig. \ref{figwork} with two bounds.
The dashed bound is obtained by comparing $\Delta \mathcal{C}$ to the inequality (\ref{DeltCB}).
}
\label{figworkbound}
\end{figure}

We now compare the bound to the enhanced concurrence for $\rho(p,\eta)$  in Eq. (\ref{example}).
Under the condition that  $T\vec{n}_{1} \perp T\vec{n}_{2}$,
the optimal two measurements on $\rho(p,\eta)$ to create the maximal quantum correlations can be found to be
$M_{\vec{n}_1} = \sigma_x $ and  $M_{\vec{n}_2}=\sigma_z$ by numerical calculation.
Then, the enhanced concurrence is given by $\Delta \mathcal{C} =[1-2 \eta +\sqrt{\eta^2 +p^2 (1-\eta^2)}]/4$.
 %Comparing the bound to the enhanced concurrence for $\rho(p,\eta)$  shown in Fig. \ref{figwork},
One can draw the quantum region in the parameter space of $(p,\eta)$, or equivalently the space of $p$ and $\Delta \mathcal{C}$.
As shown in Fig. \ref{figworkbound},
% together with the result for extractable work by the two pairs of unitaries in \cite{PRL2019}.
%The boundaries indicate that,
 the quantumness in a pure state $|\Psi\rangle$ with a larger $\eta$ is more fragile under the mixture of classical state,
although an arbitrary $|\Psi\rangle$ leads to $\bar{r}=1$.
 %%%%
To demonstrate the quantumness, the  enhanced entanglement increases with the proportion of classical state.
%And, the quantum region is narrowed in the scheme with the two pairs of non-optimal unitaries.

 %%%%

\section{summary}\label{summary}
 %%%%

We studied the preparation of quantum correlations from a system qubit and an auxiliary qubit, assisted by a Maxwell demon who obtains information of the
thermal qubit from measurements on its environment.
These processes avoid the disturbance to average state of the system by direct measurements,
and  establish the relationships between quantum steering and other correlations in the thermodynamic framework.
We derived the optimal operations between the system and the auxiliary to create the maximal mutual information, entanglement and Bell-nonlocality.
The maximums are monotonic increasing functions of the average length of the Bloch vectors in postmeasured states of the system qubit.
A critical value of the average length naturally corresponds to the necessary and sufficient condition for steerability in the case with maximally mixed marginals.
We also presented an upper bound of the average length for unsteerable environment-system correlation,
which  can be regarded as a steering-type inequality demonstrating the quantumness of the Maxwell demon.

It would be interesting to consider extensions of the current results in several directions.
On the theoretical side, one can try to derive more general relationships between the preparation of quantum correlations and the quantum steering, especially in multipartite systems and in the processes with thermodynamic cycles.
And, the result in the case with maximally mixed marginals
suggests that,
 it is possible to find
 %%%%
an operational interpretation in thermodynamic tasks
of the necessary and sufficient conditions for steerability of general two-qubit states \cite{PRA2018WuYC1,PRA2018WuYC2,PhysRevLett.122.240401}.
Experimentally,
we hope that the processes studied in this paper can be implemented in laboratories
 with the recent  techniques developed in spin systems \cite{PRL2022Ren}.

%%%%%%%%
%%%%%%%%
\begin{acknowledgments}
This work was supported by the National Natural Science Foundation of China (Grants No. 11675119, No. 11575125, and No. 11105097).
\end{acknowledgments}

%%%%

%%%%
	
\begin{appendix}

 %\section*{APPENDIX}
\section{Lemma 1}\label{Lemma1}

 \emph{Lemma 1:}
 %\begin{lemma}
Let $\{X_j\}$ be a set of  Hermitian semipositive definite operators in the $d$-dimension Hilbert space,
with the spectral decomposition $X_j= \sum_{i=1}^{d} \lambda^{(j)}_i |\phi^{(j)}_i\rangle \langle \phi^{(j)}_i |$
 and $\lambda^{(j)}_1 \geq \lambda^{(j)}_2 ...\geq \lambda^{(j)}_d$.
 For given spectrums $\{\lambda^{(j)}_i\}$,
 the maximum of $\tr (\sum_{j}X_j )^n$ $(n=2,3...)$
 occurs when the eigenstates satisfy $|\langle \phi^{(j)}_i | \phi^{(j^{\prime})}_{i^{\prime}}\rangle| =\delta_{i,i^{\prime}} $.
 \begin{proof}
 For given spectrums $\{\lambda^{(j)}_i\}$,
 the amount of $\tr (\sum_{j}X_j )^n$ depends on the cross terms in the form $\tr (X_j^{k_j} Y)$,
where $1\leq k_j \leq d-1$ and   $Y$ is a product of $X_{j^{\prime}}^{k_{j^{\prime}}}$ with $j^{\prime}\neq j$ and $\sum_{j^{\prime}} k_{j^{\prime}}+ k_j  = n$.
The operator  $Y$ is semipositive definite.
Suppose $Y=\sum_{i=1}^{d} y_i |y_i\rangle \langle y_i |$  and $y_1 \geq y_2 ...\geq y_d$.
One can directly calculate and obtain
\begin{equation}
\tr (X_j^{k_j} Y)= \sum_{ii^{\prime}} \lambda^{(j)}_i  y_{i^{\prime}} P_{ii^{\prime}},
\end{equation}
where  $P_{ii^{\prime}}=|\langle \phi^{(j)}_i | y_{i^{\prime}}\rangle|^2$ is a doubly stochastic matrix.
For given $ \lambda^{(j)}_i $ and $y_i$, $\tr (X_j^{k_j} Y)$ is maximized by $P_{ii^{\prime}}= \delta_{i,i^{\prime}}$.
For all the  cross terms and all the choices of the subscript $j$,
  the above maximization can be simultaneously  achieved by $|\langle \phi^{(j)}_i | \phi^{(j^{\prime})}_{i^{\prime}}\rangle| =\delta_{i,i^{\prime}} $.
\end{proof}

\section{Optimization of negativity}\label{MaxN}

For given eigenvalues $\{\lambda_i\}$ of two-qubit states in nonascending order, the maximal negativity is given by \cite{PRA2001MEMS}
\begin{equation}
\mathcal{N}_{\mathrm{max}} \! =\!   \max\{0, \sqrt{(\lambda_1 \! - \! \lambda_3)^2 +(\lambda_2 \! - \! \lambda_4)^2 } -\! \lambda_2 \!-\! \lambda_4 \}.
\end{equation}
It is reached by
\begin{equation}
\! \rho \!=\! \lambda_1|\psi_+\!\rangle \langle \psi_+\! | \!+\!\lambda_2|01\rangle \langle 01 | \!+\! \lambda_3|\psi_-\!\rangle \langle \psi_- \!| \!+\! \lambda_4| 10\rangle \langle 10 |.
\end{equation}
By using this result, one can derive the maximal negativity for a given the maximal eigenvalue, $\lambda_1 \in[1/4,1]$.
%Namely,

We define four lines on the plane of $(\lambda_3,\lambda_4)$ as
\begin{align}
\begin{split}\nonumber
&(a)\ \ \ \lambda_4=0,\\
&(b)\ \ \  \lambda_4=\lambda_3 ,\\
&(c)\ \ \ \lambda_4=1-\lambda_1-2\lambda_3,\\
&(d)\ \ \ \lambda_4=1-2\lambda_1-\lambda_3.\\
\end{split}
\end{align}
They are  equivalent to the four equals signs in $0\leq \lambda_4\leq \lambda_3\leq \lambda_2\leq \lambda_1$ in order.
For a fixed value of $\lambda_1$,
there are  three different situations for the physical region on the plane of $(\lambda_3,\lambda_4)$ as:
(i) a triangle defined by lines (a),  (b), and (c), when $\lambda_1 \in[1/2,1]$;
(ii) a quadrilateral defined by lines (a),  (b), (c), and (d), when $\lambda_1 \in[1/3,1/2)$;
(iii) a triangle defined by lines (b),  (c), and (d), when $\lambda_1 \in[1/4,1/3)$.
Because of the convexity of negativity, the maximum occurs on the vertices of the triangles or quadrilateral.
 Calculating the value of $\mathcal{N}_{\mathrm{max}} $ on these vertices,
 one obtains
 \begin{eqnarray}
\tilde{\mathcal{N}}_{\mathrm{max}}
 \! =\!   \max\{0,  \sqrt{10 \lambda_1^2 - 6 \lambda_1 +1} -\lambda_1, \ \ \ \ \ \ && \nonumber \\
\sqrt{2\lambda_1^2- 2\lambda_1 +1 } + \lambda_1-1\}.  &&
\end{eqnarray}
It is a monotonic increasing function of $\lambda_1$.

 The choice with $\rho_{a}=|0\rangle \langle0|$ and $U_{n|k}  \rho_{n|k} \otimes |0\rangle \langle0|  {U_{n|k}}^{\dag}  =   \frac{1+r_{n|k} }{2} |\psi_+\rangle \langle \psi_+| + \frac{1-r_{n|k} }{2}  |01\rangle \langle 01|$
 maximizes the  maximal eigenvalue of $\xi_{s a}$,
 and meanwhile leads to $\mathcal{N}(\xi_{s a})= \tilde{\mathcal{N}}_{\mathrm{max}}=\sqrt{2\lambda_1^2-2 \lambda_1 +1 } + \lambda_1-1$.
 Here $\lambda_1= \frac{1+\bar{r} }{2} \in[1/2,1]$.
Hence, this optimizes the amount of negativity.

\end{appendix}

%%%%%%%%

\bibliography{SteerDemon}

\begin{thebibliography}{46}
\expandafter\ifx\csname natexlab\endcsname\relax\def\natexlab#1{#1}\fi
\expandafter\ifx\csname bibnamefont\endcsname\relax
  \def\bibnamefont#1{#1}\fi
\expandafter\ifx\csname bibfnamefont\endcsname\relax
  \def\bibfnamefont#1{#1}\fi
\expandafter\ifx\csname citenamefont\endcsname\relax
  \def\citenamefont#1{#1}\fi
\expandafter\ifx\csname url\endcsname\relax
  \def\url#1{\texttt{#1}}\fi
\expandafter\ifx\csname urlprefix\endcsname\relax\def\urlprefix{URL }\fi
\providecommand{\bibinfo}[2]{#2}
\providecommand{\eprint}[2][]{\url{#2}}

\bibitem[{\citenamefont{Maxwell}(1911)}]{MaxwellD}
\bibinfo{author}{\bibfnamefont{J.~C.} \bibnamefont{Maxwell}},
  \emph{\bibinfo{title}{Life and Scientific Work of Peter Guthrie Tait}}
  (\bibinfo{publisher}{edited by C. G. Knott (Cambridge University Press,
  London)}, \bibinfo{year}{1911}), p. \bibinfo{pages}{213}.

\bibitem[{\citenamefont{Szil\'{a}rd}(1929)}]{Z.Phys.53.840}
\bibinfo{author}{\bibfnamefont{L.}~\bibnamefont{Szil\'{a}rd}},
  \bibinfo{journal}{Z. Phys} \textbf{\bibinfo{volume}{53}},
  \bibinfo{pages}{840} (\bibinfo{year}{1929}).

\bibitem[{\citenamefont{Gemma et~al.}(2004)\citenamefont{Gemma, Michel, and
  Mahler}}]{Book2004}
\bibinfo{author}{\bibfnamefont{J.}~\bibnamefont{Gemma}},
  \bibinfo{author}{\bibfnamefont{M.}~\bibnamefont{Michel}}, \bibnamefont{and}
  \bibinfo{author}{\bibfnamefont{G.}~\bibnamefont{Mahler}},
  \emph{\bibinfo{title}{Quantum Thermodynamics}} (\bibinfo{publisher}{Springer,
  Berlin}, \bibinfo{year}{2004}).

\bibitem[{\citenamefont{Gemmer et~al.}(2009)\citenamefont{Gemmer, Michel, and
  Mahler}}]{Book2009}
\bibinfo{author}{\bibfnamefont{J.}~\bibnamefont{Gemmer}},
  \bibinfo{author}{\bibfnamefont{M.}~\bibnamefont{Michel}}, \bibnamefont{and}
  \bibinfo{author}{\bibfnamefont{G.}~\bibnamefont{Mahler}},
  \emph{\bibinfo{title}{Quantum Thermodynamics: Emergence of Thermodynamic
  Behavior Within Composite Quantum Systems, (Lecture Notes in Physics 784)}}
  (\bibinfo{publisher}{Springer-Verlag, Heidelberg}, \bibinfo{year}{2009}).

\bibitem[{\citenamefont{Maruyama et~al.}(2009)\citenamefont{Maruyama, Nori, and
  Vedral}}]{RMP2009MaxwellD}
\bibinfo{author}{\bibfnamefont{K.}~\bibnamefont{Maruyama}},
  \bibinfo{author}{\bibfnamefont{F.}~\bibnamefont{Nori}}, \bibnamefont{and}
  \bibinfo{author}{\bibfnamefont{V.}~\bibnamefont{Vedral}},
  \bibinfo{journal}{Rev. Mod. Phys.} \textbf{\bibinfo{volume}{81}},
  \bibinfo{pages}{1} (\bibinfo{year}{2009}).

\bibitem[{\citenamefont{Zurek}(2003)}]{discorddemons}
\bibinfo{author}{\bibfnamefont{W.~H.} \bibnamefont{Zurek}},
  \bibinfo{journal}{Phys. Rev. A} \textbf{\bibinfo{volume}{67}},
  \bibinfo{pages}{012320} (\bibinfo{year}{2003}).

\bibitem[{\citenamefont{Mohammady and Anders}(2017)}]{NJP2017Szilard}
\bibinfo{author}{\bibfnamefont{M.~H.} \bibnamefont{Mohammady}}
  \bibnamefont{and} \bibinfo{author}{\bibfnamefont{J.}~\bibnamefont{Anders}},
  \bibinfo{journal}{New J. Phys.} \textbf{\bibinfo{volume}{19}},
  \bibinfo{pages}{113026} (\bibinfo{year}{2017}).

\bibitem[{\citenamefont{Elouard et~al.}(2017)\citenamefont{Elouard,
  Herrera-Mart\'{\i}, Huard, and Auff\`eves}}]{PRL2017Demon}
\bibinfo{author}{\bibfnamefont{C.}~\bibnamefont{Elouard}},
  \bibinfo{author}{\bibfnamefont{D.}~\bibnamefont{Herrera-Mart\'{\i}}},
  \bibinfo{author}{\bibfnamefont{B.}~\bibnamefont{Huard}}, \bibnamefont{and}
  \bibinfo{author}{\bibfnamefont{A.}~\bibnamefont{Auff\`eves}},
  \bibinfo{journal}{Phys. Rev. Lett.} \textbf{\bibinfo{volume}{118}},
  \bibinfo{pages}{260603} (\bibinfo{year}{2017}).

\bibitem[{\citenamefont{S\'anchez et~al.}(2019)\citenamefont{S\'anchez,
  Splettstoesser, and Whitney}}]{PRL2019NonequilibriumDemon}
\bibinfo{author}{\bibfnamefont{R.}~\bibnamefont{S\'anchez}},
  \bibinfo{author}{\bibfnamefont{J.}~\bibnamefont{Splettstoesser}},
  \bibnamefont{and} \bibinfo{author}{\bibfnamefont{R.~S.}
  \bibnamefont{Whitney}}, \bibinfo{journal}{Phys. Rev. Lett.}
  \textbf{\bibinfo{volume}{123}}, \bibinfo{pages}{216801}
  (\bibinfo{year}{2019}).

\bibitem[{\citenamefont{Beyer et~al.}(2019)\citenamefont{Beyer, Luoma, and
  Strunz}}]{PRL2019}
\bibinfo{author}{\bibfnamefont{K.}~\bibnamefont{Beyer}},
  \bibinfo{author}{\bibfnamefont{K.}~\bibnamefont{Luoma}}, \bibnamefont{and}
  \bibinfo{author}{\bibfnamefont{W.~T.} \bibnamefont{Strunz}},
  \bibinfo{journal}{Phys. Rev. Lett.} \textbf{\bibinfo{volume}{123}},
  \bibinfo{pages}{250606} (\bibinfo{year}{2019}).

\bibitem[{\citenamefont{Ji et~al.}(2022)\citenamefont{Ji, Chai, Wang, Guo,
  Rong, Shi, Ren, Wang, and Du}}]{PRL2022Ren}
\bibinfo{author}{\bibfnamefont{W.}~\bibnamefont{Ji}},
  \bibinfo{author}{\bibfnamefont{Z.}~\bibnamefont{Chai}},
  \bibinfo{author}{\bibfnamefont{M.}~\bibnamefont{Wang}},
  \bibinfo{author}{\bibfnamefont{Y.}~\bibnamefont{Guo}},
  \bibinfo{author}{\bibfnamefont{X.}~\bibnamefont{Rong}},
  \bibinfo{author}{\bibfnamefont{F.}~\bibnamefont{Shi}},
  \bibinfo{author}{\bibfnamefont{C.}~\bibnamefont{Ren}},
  \bibinfo{author}{\bibfnamefont{Y.}~\bibnamefont{Wang}}, \bibnamefont{and}
  \bibinfo{author}{\bibfnamefont{J.}~\bibnamefont{Du}}, \bibinfo{journal}{Phys.
  Rev. Lett.} \textbf{\bibinfo{volume}{128}}, \bibinfo{pages}{090602}
  (\bibinfo{year}{2022}).

\bibitem[{\citenamefont{Kim et~al.}(2011)\citenamefont{Kim, Sagawa,
  De~Liberato, and Ueda}}]{PRL2011QSzilard}
\bibinfo{author}{\bibfnamefont{S.~W.} \bibnamefont{Kim}},
  \bibinfo{author}{\bibfnamefont{T.}~\bibnamefont{Sagawa}},
  \bibinfo{author}{\bibfnamefont{S.}~\bibnamefont{De~Liberato}},
  \bibnamefont{and} \bibinfo{author}{\bibfnamefont{M.}~\bibnamefont{Ueda}},
  \bibinfo{journal}{Phys. Rev. Lett.} \textbf{\bibinfo{volume}{106}},
  \bibinfo{pages}{070401} (\bibinfo{year}{2011}).

\bibitem[{\citenamefont{Park et~al.}(2013)\citenamefont{Park, Kim, Sagawa, and
  Kim}}]{PRL2013Engine}
\bibinfo{author}{\bibfnamefont{J.~J.} \bibnamefont{Park}},
  \bibinfo{author}{\bibfnamefont{K.-H.} \bibnamefont{Kim}},
  \bibinfo{author}{\bibfnamefont{T.}~\bibnamefont{Sagawa}}, \bibnamefont{and}
  \bibinfo{author}{\bibfnamefont{S.~W.} \bibnamefont{Kim}},
  \bibinfo{journal}{Phys. Rev. Lett.} \textbf{\bibinfo{volume}{111}},
  \bibinfo{pages}{230402} (\bibinfo{year}{2013}).

\bibitem[{\citenamefont{Faist et~al.}(2015)\citenamefont{Faist, Dupuis,
  Oppenheim, and Renner}}]{NC2015}
\bibinfo{author}{\bibfnamefont{P.}~\bibnamefont{Faist}},
  \bibinfo{author}{\bibfnamefont{F.}~\bibnamefont{Dupuis}},
  \bibinfo{author}{\bibfnamefont{J.}~\bibnamefont{Oppenheim}},
  \bibnamefont{and} \bibinfo{author}{\bibfnamefont{R.}~\bibnamefont{Renner}},
  \bibinfo{journal}{Nat. Commun.} \textbf{\bibinfo{volume}{6}},
  \bibinfo{pages}{7669} (\bibinfo{year}{2015}).

\bibitem[{\citenamefont{Seah et~al.}(2020)\citenamefont{Seah, Nimmrichter, and
  Scarani}}]{PRL.124.100603}
\bibinfo{author}{\bibfnamefont{S.}~\bibnamefont{Seah}},
  \bibinfo{author}{\bibfnamefont{S.}~\bibnamefont{Nimmrichter}},
  \bibnamefont{and} \bibinfo{author}{\bibfnamefont{V.}~\bibnamefont{Scarani}},
  \bibinfo{journal}{Phys. Rev. Lett} \textbf{\bibinfo{volume}{124}},
  \bibinfo{pages}{100603} (\bibinfo{year}{2020}).

\bibitem[{\citenamefont{Huber et~al.}(2015)\citenamefont{Huber,
  Perarnau-Llobet, Hovhannisyan, Skrzypczyk, Kl\"{o}ckl, Brunner, and
  Ac\'{\i}n}}]{Huber_2015}
\bibinfo{author}{\bibfnamefont{M.}~\bibnamefont{Huber}},
  \bibinfo{author}{\bibfnamefont{M.}~\bibnamefont{Perarnau-Llobet}},
  \bibinfo{author}{\bibfnamefont{K.~V.} \bibnamefont{Hovhannisyan}},
  \bibinfo{author}{\bibfnamefont{P.}~\bibnamefont{Skrzypczyk}},
  \bibinfo{author}{\bibfnamefont{C.}~\bibnamefont{Kl\"{o}ckl}},
  \bibinfo{author}{\bibfnamefont{N.}~\bibnamefont{Brunner}}, \bibnamefont{and}
  \bibinfo{author}{\bibfnamefont{A.}~\bibnamefont{Ac\'{\i}n}},
  \bibinfo{journal}{New J. Phys.} \textbf{\bibinfo{volume}{17}},
  \bibinfo{pages}{065008} (\bibinfo{year}{2015}).

\bibitem[{\citenamefont{Guha et~al.}(2019)\citenamefont{Guha, Alimuddin, and
  Parashar}}]{PhysRevE.100.012147}
\bibinfo{author}{\bibfnamefont{T.}~\bibnamefont{Guha}},
  \bibinfo{author}{\bibfnamefont{M.}~\bibnamefont{Alimuddin}},
  \bibnamefont{and} \bibinfo{author}{\bibfnamefont{P.}~\bibnamefont{Parashar}},
  \bibinfo{journal}{Phys. Rev. E} \textbf{\bibinfo{volume}{100}},
  \bibinfo{pages}{012147} (\bibinfo{year}{2019}).

\bibitem[{\citenamefont{Oppenheim et~al.}(2002)\citenamefont{Oppenheim,
  Horodecki, Horodecki, and Horodecki}}]{PRL2002ThermodynamicalCorrelations}
\bibinfo{author}{\bibfnamefont{J.}~\bibnamefont{Oppenheim}},
  \bibinfo{author}{\bibfnamefont{M.}~\bibnamefont{Horodecki}},
  \bibinfo{author}{\bibfnamefont{P.}~\bibnamefont{Horodecki}},
  \bibnamefont{and}
  \bibinfo{author}{\bibfnamefont{R.}~\bibnamefont{Horodecki}},
  \bibinfo{journal}{Phys. Rev. Lett.} \textbf{\bibinfo{volume}{89}},
  \bibinfo{pages}{180402} (\bibinfo{year}{2002}).

\bibitem[{\citenamefont{Perarnau-Llobet
  et~al.}(2015)\citenamefont{Perarnau-Llobet, Hovhannisyan, Huber, Skrzypczyk,
  Brunner, and Ac\'{\i}n}}]{PhysRevX.5.041011}
\bibinfo{author}{\bibfnamefont{M.}~\bibnamefont{Perarnau-Llobet}},
  \bibinfo{author}{\bibfnamefont{K.~V.} \bibnamefont{Hovhannisyan}},
  \bibinfo{author}{\bibfnamefont{M.}~\bibnamefont{Huber}},
  \bibinfo{author}{\bibfnamefont{P.}~\bibnamefont{Skrzypczyk}},
  \bibinfo{author}{\bibfnamefont{N.}~\bibnamefont{Brunner}}, \bibnamefont{and}
  \bibinfo{author}{\bibfnamefont{A.}~\bibnamefont{Ac\'{\i}n}},
  \bibinfo{journal}{Phys. Rev. X} \textbf{\bibinfo{volume}{5}},
  \bibinfo{pages}{041011} (\bibinfo{year}{2015}).

\bibitem[{\citenamefont{Mukherjee et~al.}(2016)\citenamefont{Mukherjee, Roy,
  Bhattacharya, and Banik}}]{PhysRevE.93.052140}
\bibinfo{author}{\bibfnamefont{A.}~\bibnamefont{Mukherjee}},
  \bibinfo{author}{\bibfnamefont{A.}~\bibnamefont{Roy}},
  \bibinfo{author}{\bibfnamefont{S.~S.} \bibnamefont{Bhattacharya}},
  \bibnamefont{and} \bibinfo{author}{\bibfnamefont{M.}~\bibnamefont{Banik}},
  \bibinfo{journal}{Phys. Rev. E} \textbf{\bibinfo{volume}{93}},
  \bibinfo{pages}{052140} (\bibinfo{year}{2016}).

\bibitem[{\citenamefont{Alimuddin et~al.}(2019)\citenamefont{Alimuddin, Guha,
  and Parashar}}]{PhysRevA.99.052320}
\bibinfo{author}{\bibfnamefont{M.}~\bibnamefont{Alimuddin}},
  \bibinfo{author}{\bibfnamefont{T.}~\bibnamefont{Guha}}, \bibnamefont{and}
  \bibinfo{author}{\bibfnamefont{P.}~\bibnamefont{Parashar}},
  \bibinfo{journal}{Phys. Rev. A} \textbf{\bibinfo{volume}{99}},
  \bibinfo{pages}{052320} (\bibinfo{year}{2019}).

\bibitem[{\citenamefont{Francica et~al.}(2017)\citenamefont{Francica, Goold,
  Plastina, and Paternostro}}]{francica2017daemonic}
\bibinfo{author}{\bibfnamefont{G.}~\bibnamefont{Francica}},
  \bibinfo{author}{\bibfnamefont{J.}~\bibnamefont{Goold}},
  \bibinfo{author}{\bibfnamefont{F.}~\bibnamefont{Plastina}}, \bibnamefont{and}
  \bibinfo{author}{\bibfnamefont{M.}~\bibnamefont{Paternostro}},
  \bibinfo{journal}{npj Quantum Information} \textbf{\bibinfo{volume}{3}},
  \bibinfo{pages}{1} (\bibinfo{year}{2017}).

\bibitem[{\citenamefont{Manzano et~al.}(2018)\citenamefont{Manzano, Plastina,
  and Zambrini}}]{PhysRevLett.121.120602}
\bibinfo{author}{\bibfnamefont{G.}~\bibnamefont{Manzano}},
  \bibinfo{author}{\bibfnamefont{F.}~\bibnamefont{Plastina}}, \bibnamefont{and}
  \bibinfo{author}{\bibfnamefont{R.}~\bibnamefont{Zambrini}},
  \bibinfo{journal}{Phys. Rev. Lett.} \textbf{\bibinfo{volume}{121}},
  \bibinfo{pages}{120602} (\bibinfo{year}{2018}).

\bibitem[{\citenamefont{Morris et~al.}(2019)\citenamefont{Morris, Lami, and
  Adesso}}]{PhysRevLett.122.130601}
\bibinfo{author}{\bibfnamefont{B.}~\bibnamefont{Morris}},
  \bibinfo{author}{\bibfnamefont{L.}~\bibnamefont{Lami}}, \bibnamefont{and}
  \bibinfo{author}{\bibfnamefont{G.}~\bibnamefont{Adesso}},
  \bibinfo{journal}{Phys. Rev. Lett.} \textbf{\bibinfo{volume}{122}},
  \bibinfo{pages}{130601} (\bibinfo{year}{2019}).

\bibitem[{\citenamefont{Buffoni et~al.}(2019)\citenamefont{Buffoni, Solfanelli,
  Verrucchi, Cuccoli, and Campisi}}]{PRL2019MCool}
\bibinfo{author}{\bibfnamefont{L.}~\bibnamefont{Buffoni}},
  \bibinfo{author}{\bibfnamefont{A.}~\bibnamefont{Solfanelli}},
  \bibinfo{author}{\bibfnamefont{P.}~\bibnamefont{Verrucchi}},
  \bibinfo{author}{\bibfnamefont{A.}~\bibnamefont{Cuccoli}}, \bibnamefont{and}
  \bibinfo{author}{\bibfnamefont{M.}~\bibnamefont{Campisi}},
  \bibinfo{journal}{Phys. Rev. Lett.} \textbf{\bibinfo{volume}{122}},
  \bibinfo{pages}{070603} (\bibinfo{year}{2019}).

\bibitem[{\citenamefont{Wiseman et~al.}(2007)\citenamefont{Wiseman, Jones, and
  Doherty}}]{PRL2007Steering}
\bibinfo{author}{\bibfnamefont{H.~M.} \bibnamefont{Wiseman}},
  \bibinfo{author}{\bibfnamefont{S.~J.} \bibnamefont{Jones}}, \bibnamefont{and}
  \bibinfo{author}{\bibfnamefont{A.~C.} \bibnamefont{Doherty}},
  \bibinfo{journal}{Phys. Rev. Lett.} \textbf{\bibinfo{volume}{98}},
  \bibinfo{pages}{140402} (\bibinfo{year}{2007}).

\bibitem[{\citenamefont{Schr\"{o}dinger}(1935)}]{SCat}
\bibinfo{author}{\bibfnamefont{E.}~\bibnamefont{Schr\"{o}dinger}},
  \bibinfo{journal}{Proc. Cambridge Philos. Soc.}
  \textbf{\bibinfo{volume}{31}}, \bibinfo{pages}{555} (\bibinfo{year}{1935}).

\bibitem[{\citenamefont{Nielsen and Chuang}(2000)}]{Book}
\bibinfo{author}{\bibfnamefont{M.~A.} \bibnamefont{Nielsen}} \bibnamefont{and}
  \bibinfo{author}{\bibfnamefont{I.~L.} \bibnamefont{Chuang}},
  \emph{\bibinfo{title}{Quantum Computation and Quantum Information}}
  (\bibinfo{publisher}{Cambridge University Press, Cambridge},
  \bibinfo{year}{2000}).

\bibitem[{\citenamefont{Wootters}(1998)}]{Wootters98}
\bibinfo{author}{\bibfnamefont{W.~K.} \bibnamefont{Wootters}},
  \bibinfo{journal}{Phys. Rev. Lett.} \textbf{\bibinfo{volume}{80}},
  \bibinfo{pages}{2245} (\bibinfo{year}{1998}).

\bibitem[{\citenamefont{Vidal and Werner}(2002)}]{NEG1}
\bibinfo{author}{\bibfnamefont{G.}~\bibnamefont{Vidal}} \bibnamefont{and}
  \bibinfo{author}{\bibfnamefont{R.~F.} \bibnamefont{Werner}},
  \bibinfo{journal}{Phys. Rev. A} \textbf{\bibinfo{volume}{65}},
  \bibinfo{pages}{032314} (\bibinfo{year}{2002}).

\bibitem[{\citenamefont{Brunner et~al.}(2014)\citenamefont{Brunner, Cavalcanti,
  Pironio, Scarani, and Wehner}}]{RMP2014bell}
\bibinfo{author}{\bibfnamefont{N.}~\bibnamefont{Brunner}},
  \bibinfo{author}{\bibfnamefont{D.}~\bibnamefont{Cavalcanti}},
  \bibinfo{author}{\bibfnamefont{S.}~\bibnamefont{Pironio}},
  \bibinfo{author}{\bibfnamefont{V.}~\bibnamefont{Scarani}}, \bibnamefont{and}
  \bibinfo{author}{\bibfnamefont{S.}~\bibnamefont{Wehner}},
  \bibinfo{journal}{Rev. Mod. Phys.} \textbf{\bibinfo{volume}{86}},
  \bibinfo{pages}{419} (\bibinfo{year}{2014}).

\bibitem[{\citenamefont{Modi et~al.}(2012)\citenamefont{Modi, Brodutch, Cable,
  Paterek, and Vedral}}]{RMP2012Vedral}
\bibinfo{author}{\bibfnamefont{K.}~\bibnamefont{Modi}},
  \bibinfo{author}{\bibfnamefont{A.}~\bibnamefont{Brodutch}},
  \bibinfo{author}{\bibfnamefont{H.}~\bibnamefont{Cable}},
  \bibinfo{author}{\bibfnamefont{T.}~\bibnamefont{Paterek}}, \bibnamefont{and}
  \bibinfo{author}{\bibfnamefont{V.}~\bibnamefont{Vedral}},
  \bibinfo{journal}{Rev. Mod. Phys.} \textbf{\bibinfo{volume}{84}},
  \bibinfo{pages}{1655} (\bibinfo{year}{2012}).

\bibitem[{\citenamefont{Horodecki et~al.}(2009)\citenamefont{Horodecki,
  Horodecki, Horodecki, and Horodecki}}]{RevModPhys.81.865}
\bibinfo{author}{\bibfnamefont{R.}~\bibnamefont{Horodecki}},
  \bibinfo{author}{\bibfnamefont{P.}~\bibnamefont{Horodecki}},
  \bibinfo{author}{\bibfnamefont{M.}~\bibnamefont{Horodecki}},
  \bibnamefont{and}
  \bibinfo{author}{\bibfnamefont{K.}~\bibnamefont{Horodecki}},
  \bibinfo{journal}{Rev. Mod. Phys.} \textbf{\bibinfo{volume}{81}},
  \bibinfo{pages}{865} (\bibinfo{year}{2009}).

\bibitem[{\citenamefont{Ishizaka and Hiroshima}(2000)}]{PRA.62.022310}
\bibinfo{author}{\bibfnamefont{S.}~\bibnamefont{Ishizaka}} \bibnamefont{and}
  \bibinfo{author}{\bibfnamefont{T.}~\bibnamefont{Hiroshima}},
  \bibinfo{journal}{Phys. Rev. A} \textbf{\bibinfo{volume}{62}},
  \bibinfo{pages}{022310} (\bibinfo{year}{2000}).

\bibitem[{\citenamefont{Verstraete
  et~al.}(2001{\natexlab{a}})\citenamefont{Verstraete, Audenaert, and
  De~Moor}}]{PRA2001MEMS}
\bibinfo{author}{\bibfnamefont{F.}~\bibnamefont{Verstraete}},
  \bibinfo{author}{\bibfnamefont{K.}~\bibnamefont{Audenaert}},
  \bibnamefont{and} \bibinfo{author}{\bibfnamefont{B.}~\bibnamefont{De~Moor}},
  \bibinfo{journal}{Phys. Rev. A} \textbf{\bibinfo{volume}{64}},
  \bibinfo{pages}{012316} (\bibinfo{year}{2001}{\natexlab{a}}).

\bibitem[{\citenamefont{Verstraete
  et~al.}(2001{\natexlab{b}})\citenamefont{Verstraete, Audenaert, Dehaene, and
  Moor}}]{verstraete2001comparison}
\bibinfo{author}{\bibfnamefont{F.}~\bibnamefont{Verstraete}},
  \bibinfo{author}{\bibfnamefont{K.}~\bibnamefont{Audenaert}},
  \bibinfo{author}{\bibfnamefont{J.}~\bibnamefont{Dehaene}}, \bibnamefont{and}
  \bibinfo{author}{\bibfnamefont{B.}~\bibnamefont{Moor}}, \bibinfo{journal}{J.
  Phys. A} \textbf{\bibinfo{volume}{34}}, \bibinfo{pages}{10327}
  (\bibinfo{year}{2001}{\natexlab{b}}).

\bibitem[{\citenamefont{Horodecki et~al.}(1995)\citenamefont{Horodecki,
  Horodecki, and Horodecki}}]{horodecki1995violating}
\bibinfo{author}{\bibfnamefont{R.}~\bibnamefont{Horodecki}},
  \bibinfo{author}{\bibfnamefont{P.}~\bibnamefont{Horodecki}},
  \bibnamefont{and}
  \bibinfo{author}{\bibfnamefont{M.}~\bibnamefont{Horodecki}},
  \bibinfo{journal}{Phys. Lett. A} \textbf{\bibinfo{volume}{200}},
  \bibinfo{pages}{340} (\bibinfo{year}{1995}).

\bibitem[{\citenamefont{Su et~al.}(2016)\citenamefont{Su, Ren, Chen, Zhang, Wu,
  Xu, Gu, Vinjanampathy, and Kwek}}]{ZhangPRA2016}
\bibinfo{author}{\bibfnamefont{H.-Y.} \bibnamefont{Su}},
  \bibinfo{author}{\bibfnamefont{C.}~\bibnamefont{Ren}},
  \bibinfo{author}{\bibfnamefont{J.-L.} \bibnamefont{Chen}},
  \bibinfo{author}{\bibfnamefont{F.-L.} \bibnamefont{Zhang}},
  \bibinfo{author}{\bibfnamefont{C.}~\bibnamefont{Wu}},
  \bibinfo{author}{\bibfnamefont{Z.-P.} \bibnamefont{Xu}},
  \bibinfo{author}{\bibfnamefont{M.}~\bibnamefont{Gu}},
  \bibinfo{author}{\bibfnamefont{S.}~\bibnamefont{Vinjanampathy}},
  \bibnamefont{and} \bibinfo{author}{\bibfnamefont{L.~C.} \bibnamefont{Kwek}},
  \bibinfo{journal}{Phys. Rev. A} \textbf{\bibinfo{volume}{93}},
  \bibinfo{pages}{022110} (\bibinfo{year}{2016}).

\bibitem[{\citenamefont{Horodecki and
  Horodecki}(1996)}]{horodecki1996information}
\bibinfo{author}{\bibfnamefont{R.}~\bibnamefont{Horodecki}} \bibnamefont{and}
  \bibinfo{author}{\bibfnamefont{M.}~\bibnamefont{Horodecki}},
  \bibinfo{journal}{Phys. Rev. A} \textbf{\bibinfo{volume}{54}},
  \bibinfo{pages}{1838} (\bibinfo{year}{1996}).

\bibitem[{\citenamefont{Daki{\'c} et~al.}(2010)\citenamefont{Daki{\'c}, Vedral,
  and Brukner}}]{dakic2010necessary}
\bibinfo{author}{\bibfnamefont{B.}~\bibnamefont{Daki{\'c}}},
  \bibinfo{author}{\bibfnamefont{V.}~\bibnamefont{Vedral}}, \bibnamefont{and}
  \bibinfo{author}{\bibfnamefont{{\v{C}}.}~\bibnamefont{Brukner}},
  \bibinfo{journal}{Phys. Rev. Lett.} \textbf{\bibinfo{volume}{105}},
  \bibinfo{pages}{190502} (\bibinfo{year}{2010}).

\bibitem[{\citenamefont{Zhang and Zhang}(2019)}]{PRA.99.062314}
\bibinfo{author}{\bibfnamefont{F.-L.} \bibnamefont{Zhang}} \bibnamefont{and}
  \bibinfo{author}{\bibfnamefont{Y.-Y.} \bibnamefont{Zhang}},
  \bibinfo{journal}{Phys. Rev. A} \textbf{\bibinfo{volume}{99}},
  \bibinfo{pages}{062314} (\bibinfo{year}{2019}).

\bibitem[{\citenamefont{Jevtic et~al.}(2015)\citenamefont{Jevtic, Hall,
  Anderson, Zwierz, and Wiseman}}]{JOSAB2015Steering}
\bibinfo{author}{\bibfnamefont{S.}~\bibnamefont{Jevtic}},
  \bibinfo{author}{\bibfnamefont{M.~J.} \bibnamefont{Hall}},
  \bibinfo{author}{\bibfnamefont{M.~R.} \bibnamefont{Anderson}},
  \bibinfo{author}{\bibfnamefont{M.}~\bibnamefont{Zwierz}}, \bibnamefont{and}
  \bibinfo{author}{\bibfnamefont{H.~M.} \bibnamefont{Wiseman}},
  \bibinfo{journal}{J. Opt. Soc. Am. B} \textbf{\bibinfo{volume}{32}},
  \bibinfo{pages}{A40} (\bibinfo{year}{2015}).

\bibitem[{\citenamefont{Nguyen and Vu}(2016)}]{EPL2016Tstate}
\bibinfo{author}{\bibfnamefont{H.~C.} \bibnamefont{Nguyen}} \bibnamefont{and}
  \bibinfo{author}{\bibfnamefont{T.}~\bibnamefont{Vu}},
  \bibinfo{journal}{Europhys. Lett.} \textbf{\bibinfo{volume}{115}},
  \bibinfo{pages}{10003} (\bibinfo{year}{2016}).

\bibitem[{\citenamefont{Yu et~al.}(2018{\natexlab{a}})\citenamefont{Yu, Jia,
  Wu, and Guo}}]{PRA2018WuYC1}
\bibinfo{author}{\bibfnamefont{B.-C.} \bibnamefont{Yu}},
  \bibinfo{author}{\bibfnamefont{Z.-A.} \bibnamefont{Jia}},
  \bibinfo{author}{\bibfnamefont{Y.-C.} \bibnamefont{Wu}}, \bibnamefont{and}
  \bibinfo{author}{\bibfnamefont{G.-C.} \bibnamefont{Guo}},
  \bibinfo{journal}{Phys. Rev. A} \textbf{\bibinfo{volume}{97}},
  \bibinfo{pages}{012130} (\bibinfo{year}{2018}{\natexlab{a}}).

\bibitem[{\citenamefont{Yu et~al.}(2018{\natexlab{b}})\citenamefont{Yu, Jia,
  Wu, and Guo}}]{PRA2018WuYC2}
\bibinfo{author}{\bibfnamefont{B.-C.} \bibnamefont{Yu}},
  \bibinfo{author}{\bibfnamefont{Z.-A.} \bibnamefont{Jia}},
  \bibinfo{author}{\bibfnamefont{Y.-C.} \bibnamefont{Wu}}, \bibnamefont{and}
  \bibinfo{author}{\bibfnamefont{G.-C.} \bibnamefont{Guo}},
  \bibinfo{journal}{Phys. Rev. A} \textbf{\bibinfo{volume}{98}},
  \bibinfo{pages}{052345} (\bibinfo{year}{2018}{\natexlab{b}}).

\bibitem[{\citenamefont{Nguyen et~al.}(2019)\citenamefont{Nguyen, Nguyen, and
  G\"uhne}}]{PhysRevLett.122.240401}
\bibinfo{author}{\bibfnamefont{H.~C.} \bibnamefont{Nguyen}},
  \bibinfo{author}{\bibfnamefont{H.-V.} \bibnamefont{Nguyen}},
  \bibnamefont{and} \bibinfo{author}{\bibfnamefont{O.}~\bibnamefont{G\"uhne}},
  \bibinfo{journal}{Phys. Rev. Lett.} \textbf{\bibinfo{volume}{122}},
  \bibinfo{pages}{240401} (\bibinfo{year}{2019}).

\end{thebibliography}
 %%%%
	
\end{document}